\documentclass[a4paper,11pt]{article}
\pdfoutput=1 

\usepackage{jinstpub} 

\title{CHEC: A Compact High Energy Camera for the Cherenkov Telescope Array}

\author{Richard White}
\affiliation{Max-Planck-Institut f\"{u}r Kernphysik, P.O. Box 103980, 69029 Heidelberg, Germany}
\emailAdd{richard.white@mpi-hd.mpg.de}

\collaboration[c]{for the CTA GCT project \\ http://www.cta-observatory.org}


\arxivnumber{} 

\abstract{
The Cherenkov Telescope Array will provide unprecedented sensitivity and angular resolution to gamma rays across orders of magnitude in energy. Above 1~TeV up to around 300~TeV an array of Small-Sized Telescopes (SSTs) will cover several kilometres on the ground. The Compact High-Energy Camera (CHEC) is a proposed option for the camera of the SSTs. CHEC contains 2048 pixels of physical size about 6~mm~$\times$~6~mm, leading to a field of view of over 8 degrees. Electronics based on custom ASICs (TARGET) and FPGAs sample incoming signals at a gigasample per second and provide a flexible triggering scheme. Waveforms for every pixel in every event are read out without loss at over 600 events per second. A telescope prototype in Meudon, Paris, saw first Cherenkov light from air showers in late 2015, using the first CHEC prototype. Research and development for CHEC is currently focussed on taking advantage of the latest generation of silicon photomultipliers (SiPMs). 
}

\begin{document}
\maketitle
\flushbottom

\section{Introduction}
\label{intro}

\begin{figure}[t]
	\centering
	\makebox[1\columnwidth][c]{\includegraphics[trim=2.5cm 3cm 2cm 4cm,
		clip=true, width=1\textwidth]{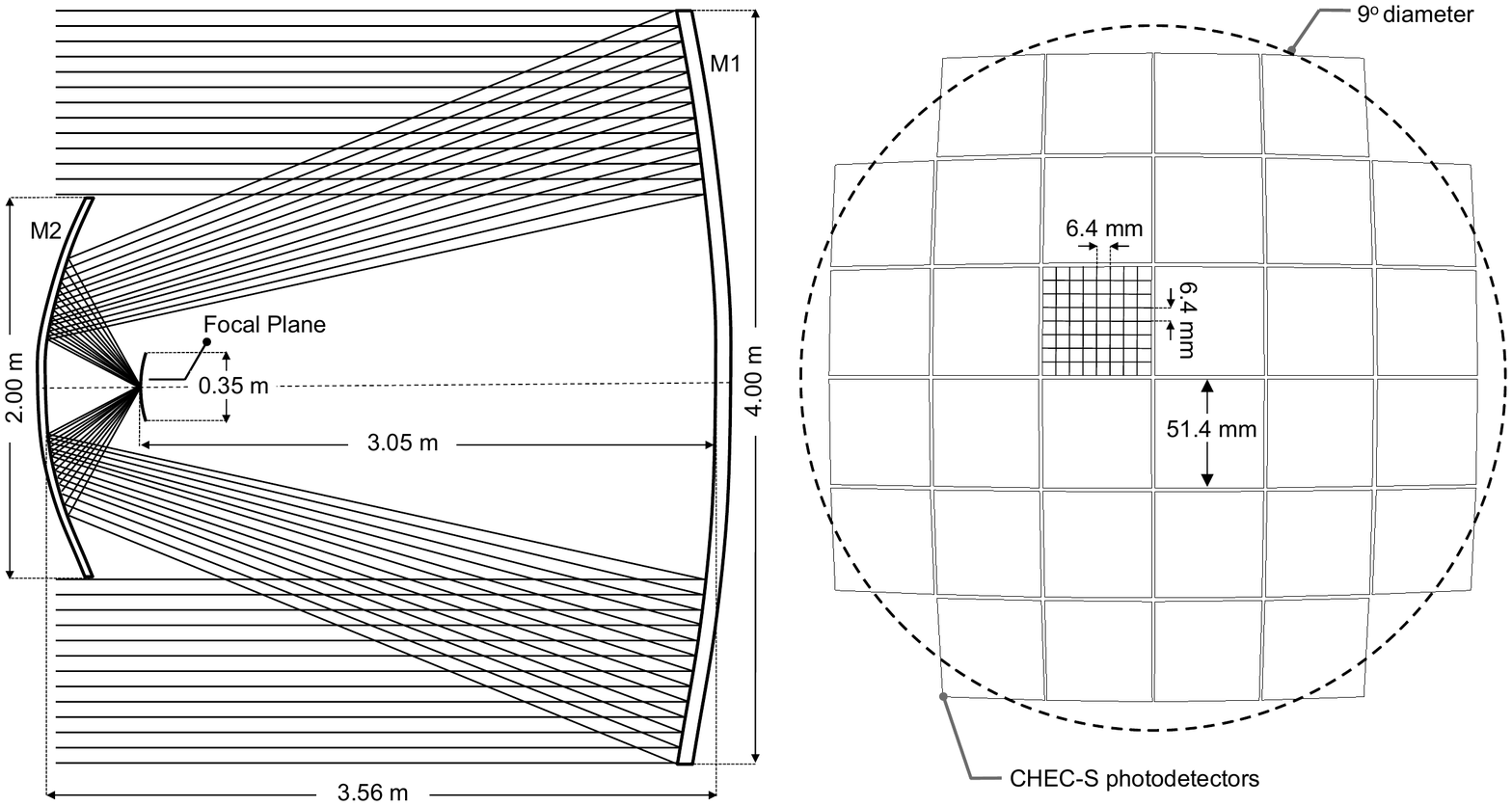}}
	\caption{A schematic showing an overview of the dual-mirror telescope geometry (left), and CHEC geometry as viewed towards the camera face (right).}
	\label{fig:dualmirror}
\end{figure}

The Cherenkov Telescope Array (CTA) is a forthcoming ground-based gamma-ray observatory designed to reach an unprecedented level of sensitivity over four orders of magnitude in energy~\cite{cta}. To achieve this CTA relies on the imaging atmospheric Cherenkov technique and the use multiple telescope sizes: the Large, Medium and Small-sized Telescopes (LSTs, MSTs, and SSTs). The observatory will host $\sim$70 SSTs~\cite{sst-icrc} in the Southern Hemisphere, providing sensitivity in an energy range from about 1 to 300 TeV, the upper end of CTA's energy reach, which extends down to 20 GeV, and an angular resolution unmatched by any instrument above X-ray energies. The Gamma-ray Cherenkov Telescope (GCT) is one option proposed for the SSTs~\cite{gct-gamma16,gct-tel}. The GCT telescope is a dual-mirror Schwarzschild-Couder~\citep{Schwarzschild1905, Couder1926, Vassiliev:2007fk} design with a primary mirror of diameter 4~m, a secondary mirror of diameter 2~m and a focal length of 2.3~m. The focal surface is spherical, with a radius of curvature 1.0~m~\cite{robast} as shown in Figure~\ref{fig:dualmirror}(left). The GCT will be equipped with the Compact High-Energy Camera (CHEC), which is also compatible with the SST design proposed by the ASTRI groups of CTA~\cite{astri}. CHEC is designed to record flashes of Cherenkov light lasting from a few to a few tens of nanoseconds, with typical image widths and lengths of $\sim$0.2$^{\circ} \times $1.0$^{\circ}$ and is designed to provide a high-reliability, high-data-quality, low-cost instrument to CTA. The desired minimum image width of $\sim$0.2$^{\circ}$ is resolvable given the telescope optics with pixels of physical dimensions of 6 to 7~mm. The dual-mirror optics maintain a point spread function (PSF) below 6~mm up to field angles of 4.5$^{\circ}$. A field of view (FoV) of 8-9$^{\circ}$ (required to capture high-energy and off-axis events) can then be covered with a camera composed of 2048 pixels covering a diameter of approximately 0.35~m as shown in Figure~\ref{fig:dualmirror}(right). Such a geometry leads to the use of commercially available photosensor arrays, significantly reducing the complexity and cost of the camera. Multi-anode photomultipliers (MAPMs) and silicon photomultipliers (SiPMs) are under investigation via the development of two prototypes CHEC-M and CHEC-S respectively.

\section{Camera Design}
\label{design}

\begin{figure}[t]
	\centering
	\makebox[1\columnwidth][c]{\includegraphics[trim=2cm 2.5cm 2.3cm 0.9cm,
		clip=true, width=1\textwidth]{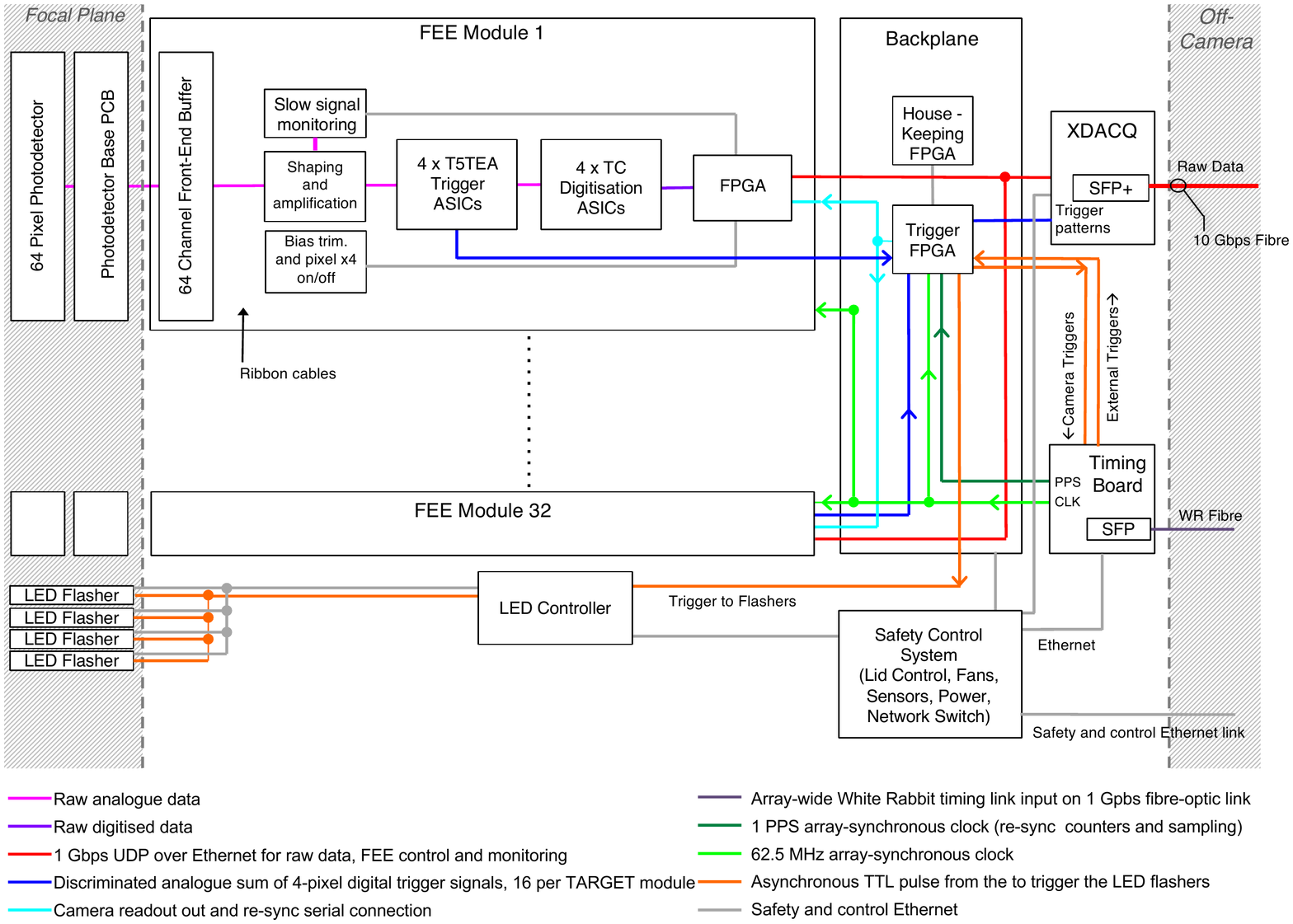}}
	\caption{A schematic showing the logical elements of CHEC,
		the communication between those elements, the raw data flow
		through the camera, the trigger architecture and the clock
		distribution scheme. Power distribution is excluded for
		simplicity.}
	\label{fig:arch}
\end{figure}

\begin{figure}[t]
	\centering
	\resizebox{1\columnwidth}{!}{\includegraphics[trim=2.5cm 1cm 2.3cm 5.8cm, clip=true]{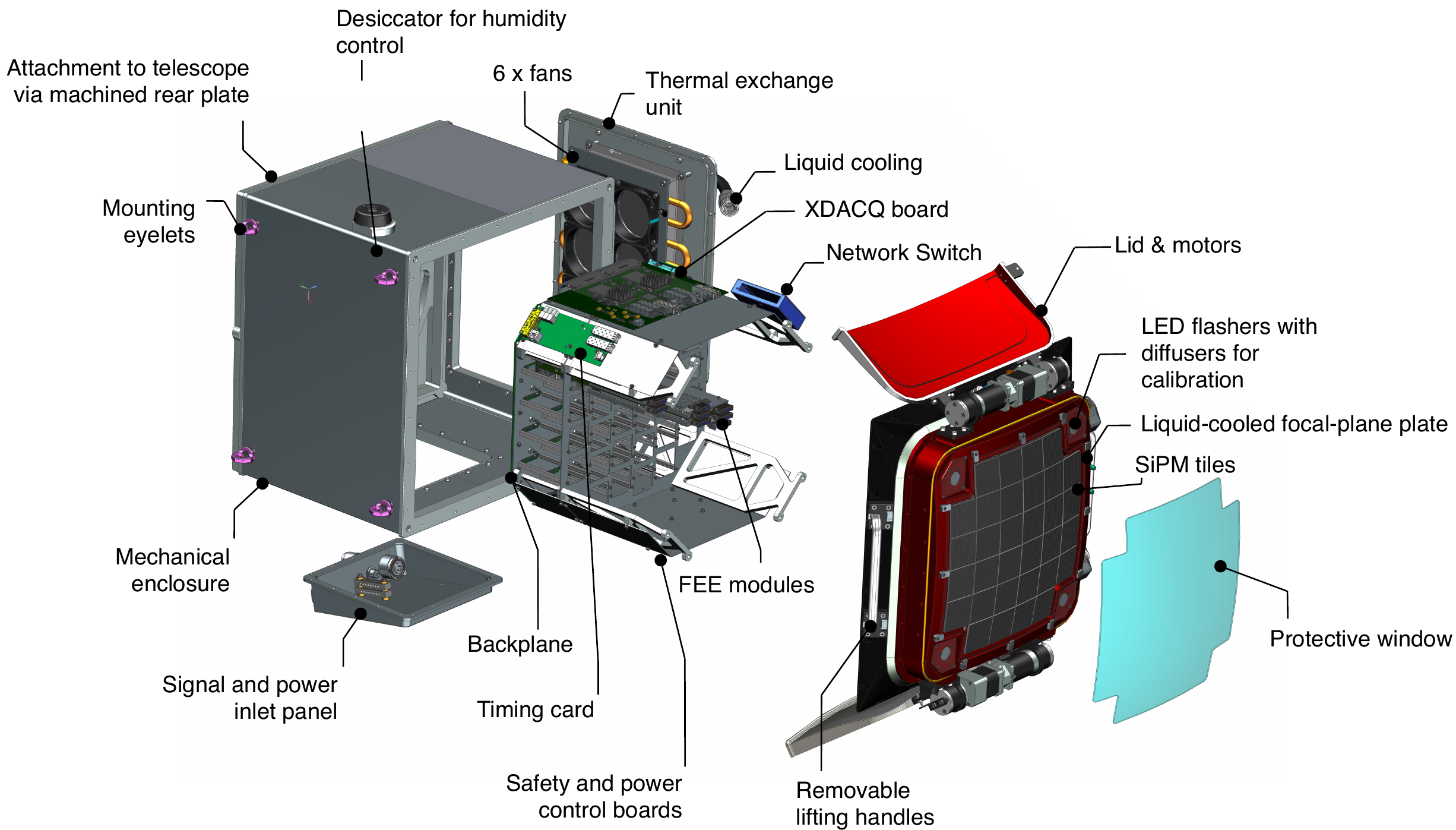}}
	\caption{The CHEC-S CAD model with the key elements highlighted.}
	\label{fig:checs-expand}
\end{figure}

CHEC consists of 2048 pixels grouped into 32 camera modules. Each module contains 64 pixels nominally $\sim$6$\times$6~mm$^2$ in size and arranged in the focal plane to approximate the required radius of curvature resulting from the telescope optics. Internally camera modules connect to an intelligent backplane that provides camera triggering and routes data to a data acquisition (DACQ) board. The architecture of the CHEC internal elements is shown in Figure~\ref{fig:arch} whilst Figure~\ref{fig:checs-expand} shows an exploded CAD image indicating the key camera components. 

\subsection{Photodetectors}

\begin{figure}[t]
	\centering
	\resizebox{0.9\columnwidth}{!}{\includegraphics[trim=2cm 10cm 2cm 0cm, clip=true]{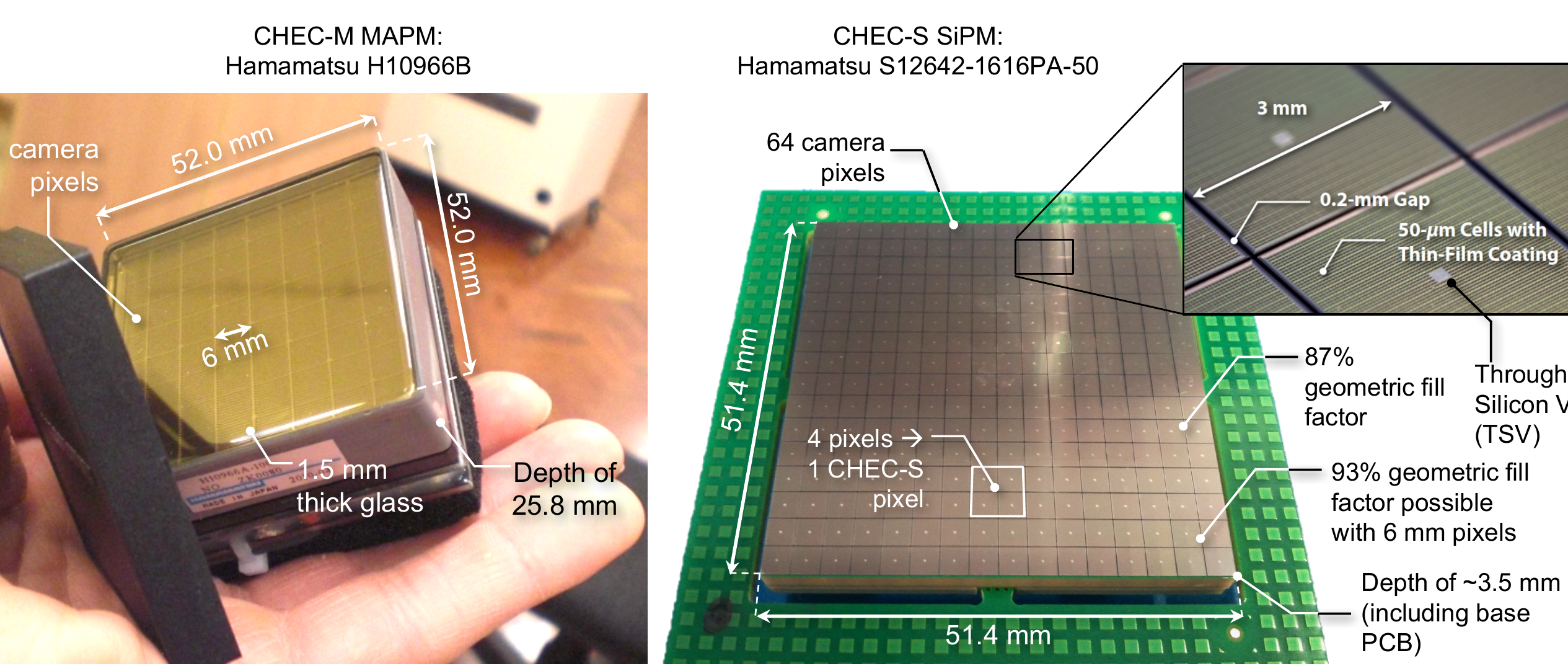}}
	\caption{The photosensors used in the CHEC-M and CHEC-S prototype cameras.}
	\label{fig:sensors}
\end{figure}

Figure~\ref{fig:sensors} shows the photosensors used in CHEC-M and CHEC-S. CHEC-M is based on Hamamatsu H10966B MAPMs, whilst CHEC-S utilises Hamamatsu S12642-1616PA-50 SiPM tiles (refer to ~\cite{checm-icrc} for details). Each CHEC-S SiPM tile contains 256 3$\times$3~mm$^2$ pixels, combined in groups of four on a bias board directly mounted to the SiPM to provide the desired camera pixel size. See Section~\ref{future} for information about the final choice of photosensor. 

\subsection{Front-End Electronics}

\begin{figure}[t]
	\centering
	\resizebox{1\columnwidth}{!}{\includegraphics[trim=2cm 4.5cm 2cm 1cm, clip=true]{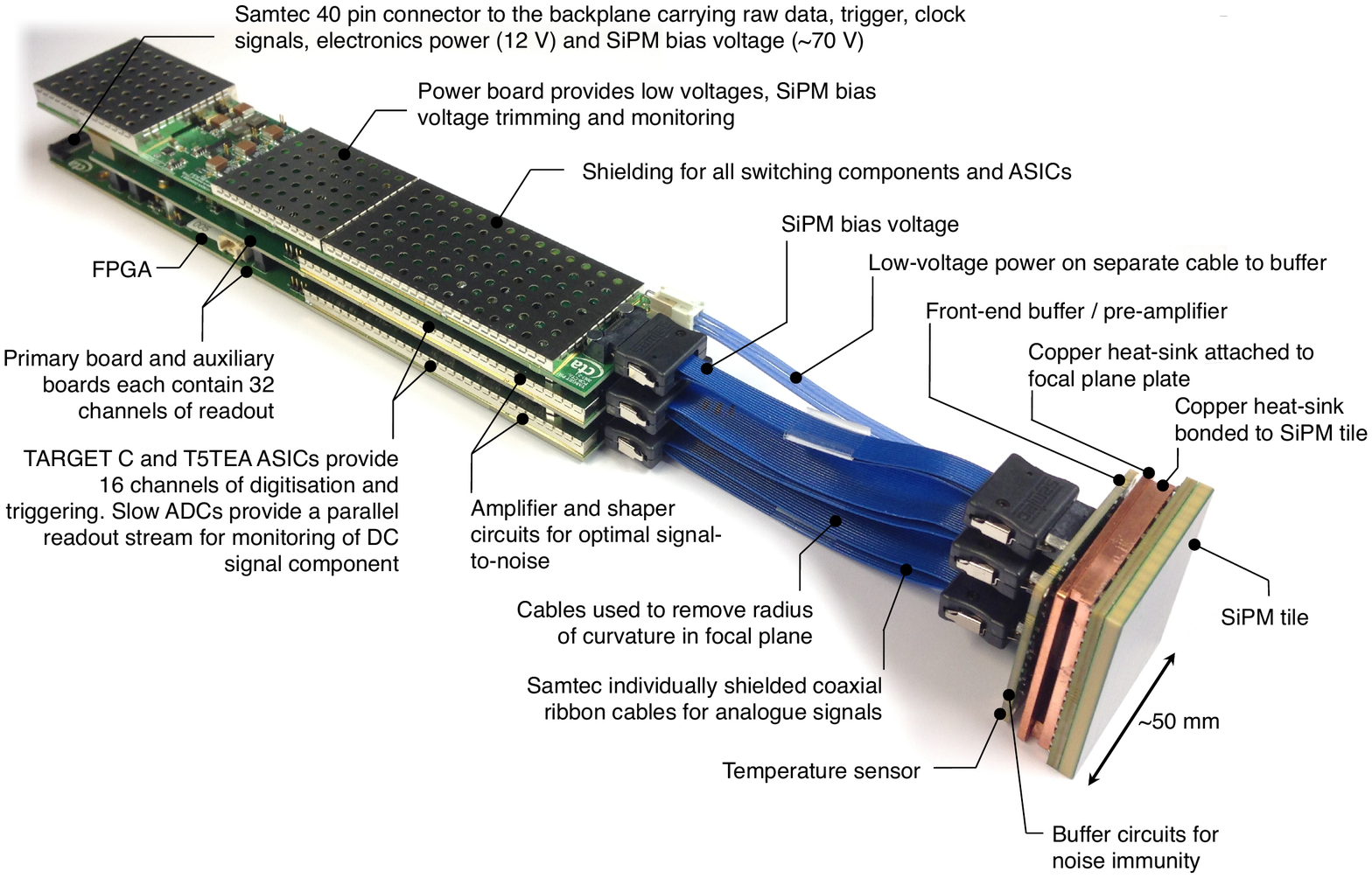}}
	\caption{The CHEC-S SiPM tile and FEE module. The final FEE module is likely to be very similar to that shown, whereas the SiPM will see an upgrade to the latest technology (see Section~\ref{future}).}
	\label{fig:tm}
\end{figure}

Each photodetector is connected to a Front-End Electronics (FEE) module based on TARGET ASICs~\cite{target-gamma16} that provides waveform digitisation and a first-level camera trigger (implemented as a binary output from the analogue sum of four neighbouring pixels passing a threshold).The CHEC-S FEE module is shown in Figure~\ref{fig:tm}. The FEE module includes a pre-amplifier board that connects to the photosensor and provides noise immunity for signals sent over individually shielded ribbon cables to the sampling and trigger boards. The optimal pulse width for triggering is around 5 to 10~ns. Any wider and night-sky-background (NSB) photons limit the performance of the analogue sum trigger, any narrower and the time gradient of the Cherenkov images prevents the coincidence of neighbouring pixel signals, the analogue sum of which is required to reach the trigger threshold. To achieve the desired pulse shape in CHEC-M the MAPM pulses are widened via the preamplifier directly behind the MAPM. The SiPM pulses in CHEC-S require shortening, which is achieved via a zero-pole shaping circuit on the FEE modules. 

The CHEC-M FEE modules are based on TARGET 5 ASICs, 16-channel devices combining digitisation and triggering functionalities~\cite{t5}. Since the development of the CHEC-M prototype the TARGET ASIC has undergone significant development. The latest generation of TARGET ASICs are used in the CHEC-S FEE modules. Four 16-channel TARGET C ASICs provide sampling whilst four 16-channel T5TEA ASICs are used for triggering. The TARGET C ASIC is a 12-bit device that provides an effective dynamic range of 1 to $\sim$500~pe for CHEC. The recovery of larger signals off line is possible due to the waveform digitisation. The sampling rate of TARGET C is tunable, but nominally set to 1~GSa/s for CHEC. This ASIC contains a 64~ns deep analogue sampling array followed by a storage array with a maximum depth of 16384~ns. To minimise calibration (required for every cell used), the storage array depth is configured to 4096~ns for CHEC-S. The size of the readout window digitised from the storage array is settable in 32~ns blocks with a position accurate to 1~ns and is set to 96~ns for CHEC to capture high-energy, off axis events as they transit the focal plane. A slow-signal digitisation chain providing a per-pixel measurement of the DC light level in the photosensors is included in the FEE modules. This may be used to track the pointing of the telescope via stars during normal operation. The TARGET module requires $\sim$70~V for the SiPM bias voltage, which is then trimmed to a precise value per four camera pixels on-board and 12~V for all other use. Module control and raw data output is via UDP over a 1~Gbps Ethernet link at the rear of the modules. An FPGA on-board each FEE module is used to configure the ASICs and other module components, to read-out raw data from the ASICs, and to package and buffer raw data for output from the module. 

\subsection{Back-End Electronics}

The 32 camera FEE modules are connected to a backplane that provides the interface for power, clock, trigger and data. It forms a nanosecond-accurate camera trigger decision by combing signals from all FEE modules in a single FPGA. The FPGA used for triggering is a Xilinx Virtex 6, which accepts all 512 first-level trigger lines from the FEE modules and implements a camera-level trigger algorithm (nominally set to require a coincidence between two neighbouring FEE trigger patches). Following a camera trigger a serial message containing a unique event identifier is sent to the FEE modules to retrieve data from the sampling ASICs at the appropriate position in their memory. Data and communication links to the FEE modules are routed over the backplane to the DACQ board and then off-camera via fibre-optic link. Two prototype DACQ boards were used in CHEC-M providing four 1~Gbps links. In the final system a new DACQ board (referred to as the XDACQ board, and currently being tested) will provide a single 10~Gbps to the camera. An array-wide White Rabbit system (\cite{wr}) connected to a timing board inside the camera provides absolute timing. A safety board controls power to camera components based on monitored environmental conditions. 

\subsection{Calibration and Environmental Control}

\begin{figure}[t]
	\centering
	\resizebox{1\columnwidth}{!}{\includegraphics[trim=1cm 5cm 1cm 1cm, clip=true]{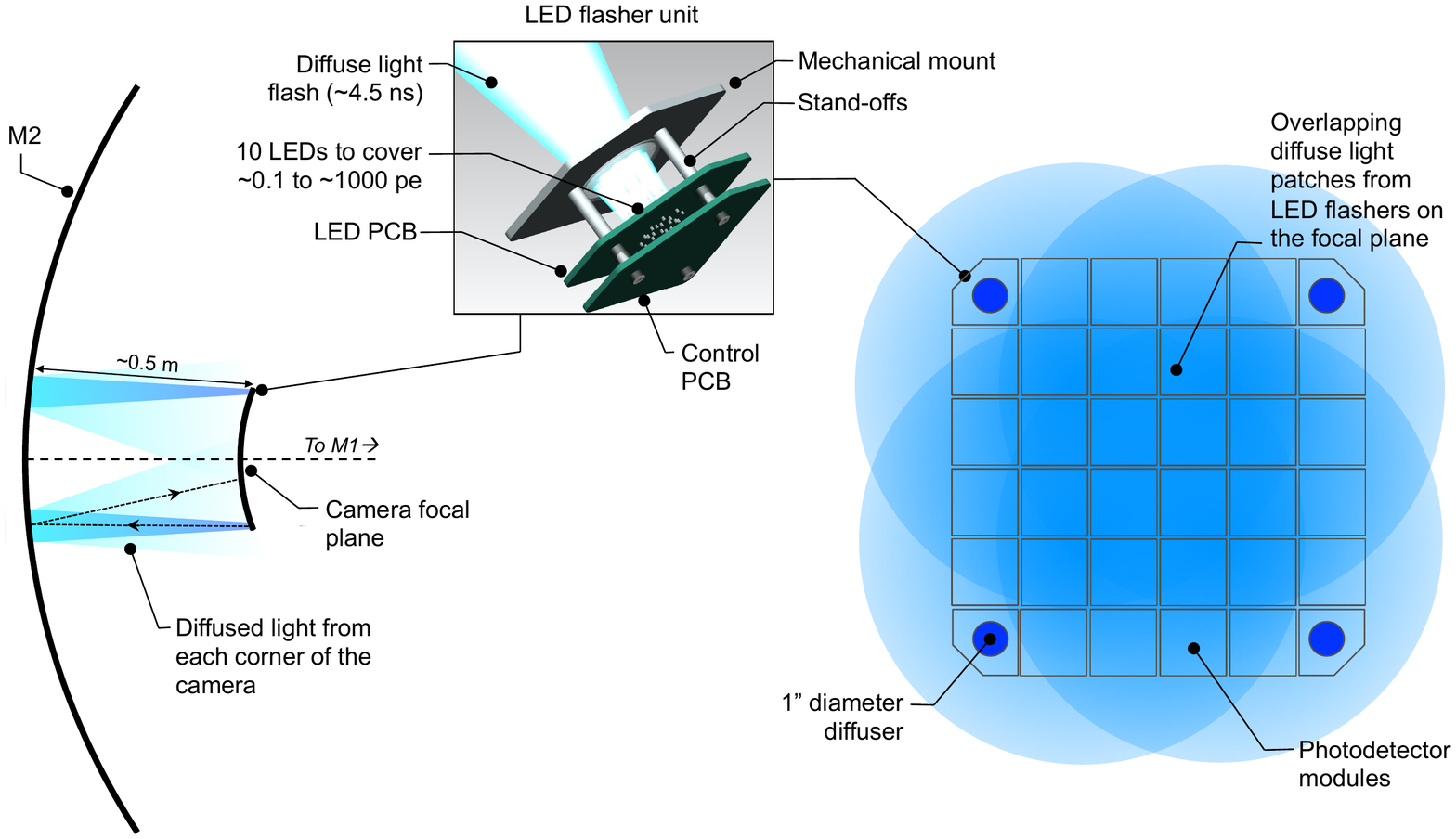}}
	\caption{The calibration flasher geometry for CHEC, with CAD model of the LED flasher unit inset.}
	\label{fig:led}
\end{figure}

Each corner of the camera is equipped with an LED flasher unit to provide calibration via reflection from the secondary mirror as shown in Figure~\ref{fig:led}. The LED flasher units each contain ten LEDs, configured in a set of patterns to provide illumination over a wide range of intensities~\cite{led-icrc2015}. An external lid system provides protection from the elements. Thermal control of the camera is via an external chiller mounted on the telescope. Chilled liquid is circulated through the camera focal plane plate (via hollow ribs) and a thermal exchange unit on the camera body. Six fans internal to the camera circulate the resulting cooled air. Cooling the focal plane plate allows the SiPM temperature to be maintained at a level desirable to stabilise the gain. The camera is hermetically sealed and a breather-desiccator is used to maintain an acceptable level of humidity and atmospheric pressure within the instrument.

\section{Prototype Commissioning and Performance}
\label{proto}

\subsection{CHEC-M}
\label{checm}

\begin{figure}[t]
	\centering
	\includegraphics[angle=0,trim=4.2cm 2cm 4.5cm 1.5cm,
	clip=true,width=0.48\textwidth]{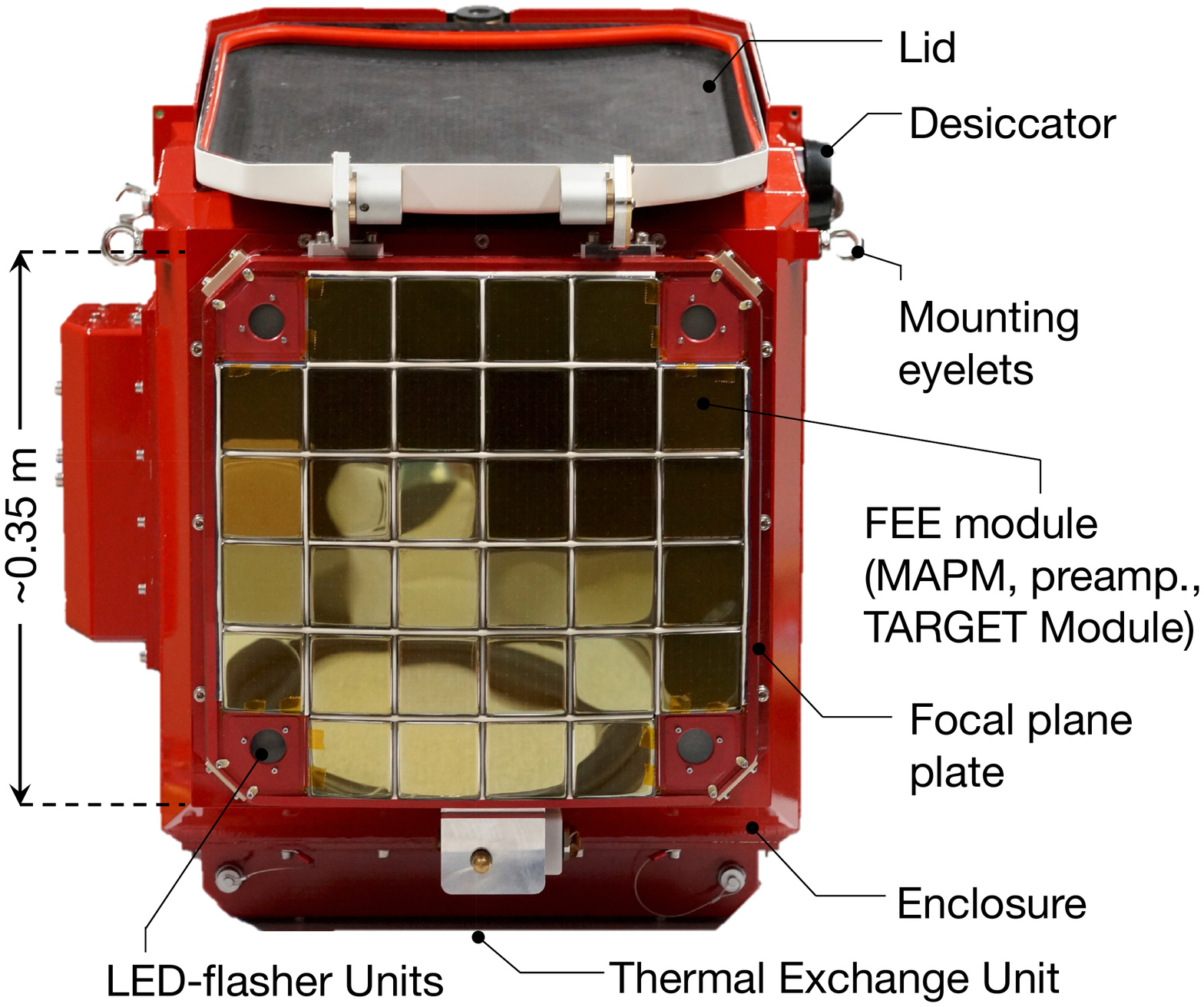}
	\includegraphics[angle=0,trim=4.5cm 3cm 2cm 0cm, clip=true,width=0.48\textwidth]{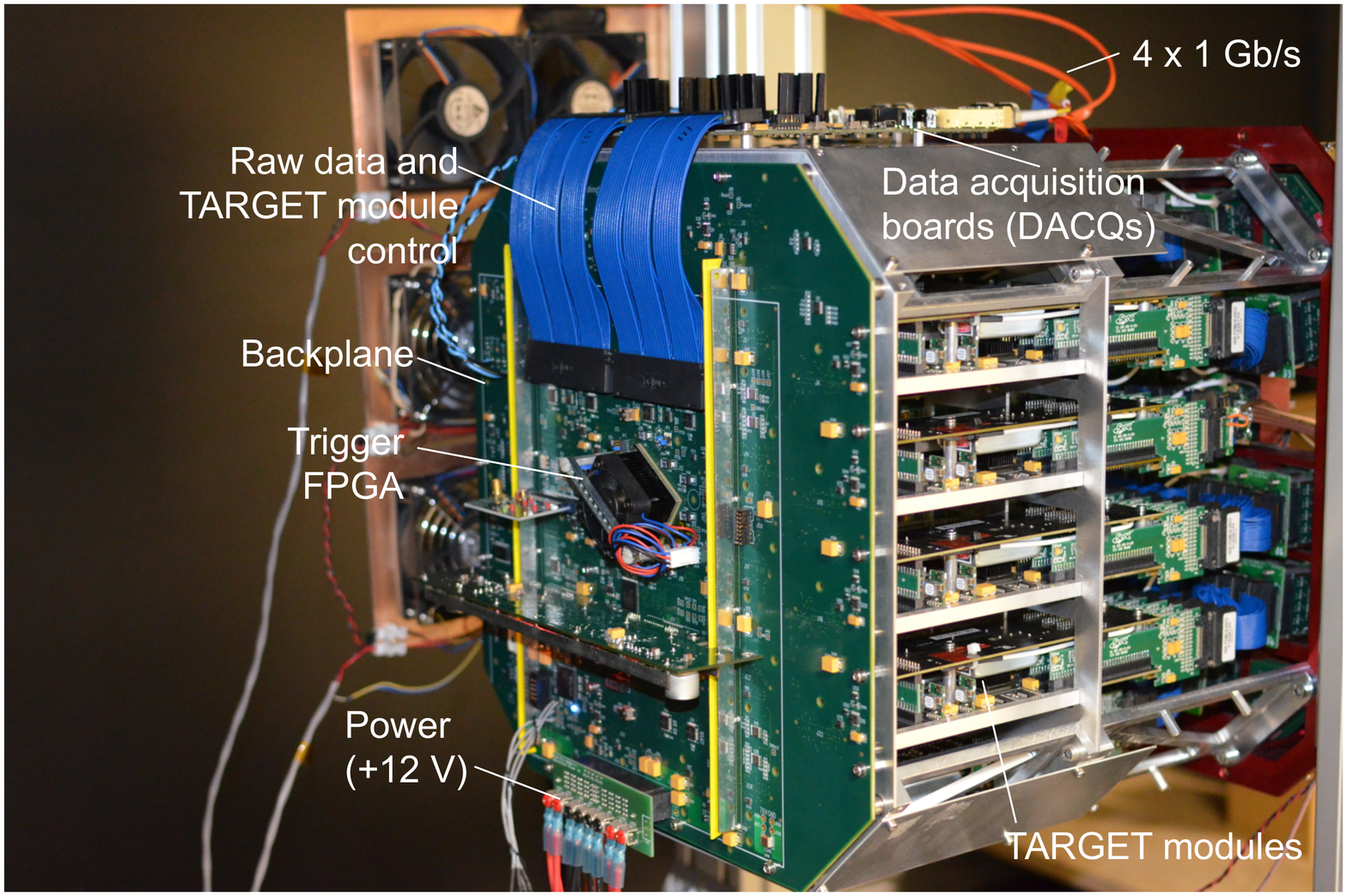}
	\caption[]{Left: The CHEC-M prototype camera, with major elements indicated. Right: A photograph of the CHEC-M camera without the external enclosure in place, taken from the back to highlight the backplane. The TARGET modules can be seen inserted into the internal rack mechanics, whilst the DACQ boards can be seen at the top of the rack, attached to the backplane via two large Samtec ribbon cables.}
	\label{fig:checm}
\end{figure}

Figure~\ref{fig:checm} shows the fully assembled CHEC-M prototype camera (left), and internal electronics (right). The CHEC-M components and fully-assembled camera have been extensively tested. Lessons learnt from CHEC-M have been incorporated into CHEC-S. Whilst a detailed paper is in preparation, here a brief summary is given. 

\paragraph{Laboratory Testing}

The single photoelectron (pe) peak can be resolved for all pixels of CHEC-M when the MAPMs are operated at the manufacturer's highest recommended voltage (1100~V). As the MAPMs are operated at a lower gain during normal operation extrapolation is required from the 1100~V single pe measurements to gain-match the camera. Each MAPM accepts only a single HV supply for all 64 pixels. This results in an unavoidable spread in gain of $\sim$20\% RMS. Whilst the gain of individual pixels may be calibrated off-line for event reconstruction, on-line CHEC-M may not be gain matched to better than this spread, an important factor in the performance of the camera trigger. The dual-mirror optical system results in light at angles of up to 70$^\circ$ impinging on the focal plane, and therefore the angular dependence of the detector efficiency is important. For MAPMs a significant loss (20-30\%) at the largest field angles is observed. The overall angular-averaged detection efficiency of a CHEC-M pixel, including camera dead space is around 14.5\%. 

A single pe pulse from the MAPM, with typical amplitude of 0.8~mV results in a pulse of typical peak voltage 2.4~mV and FWHM 5.5$\pm$1~ns as measured across all 2048 pixels at the pre-amplifier output. The maximum pre-amplifier output pulse height of $\sim$1.2~V is matched to the maximum input voltage of the TARGET 5 ASICs and corresponds to $\sim$500~pe. The output of the H10966 is linear to within 20\% at 1000~pe, and the dynamic range of CHEC-M is therefore limited by the digitisation range of the TARGET 5 ASICs. Charge resolution above saturation is possible by fitting the saturated waveforms. One ADC count in CHEC-M corresponds to approximately 0.3~mV or 0.13~pe, quantisation errors are therefore always much smaller than poisson fluctuations. The electronic noise of the FEE module system is $\sim$1~mV (or $\sim$0.5~pe) RMS. Timing measurements indicate a pulse arrival time of $\pm$0.9~ns relative to the camera trigger, within the CTA requirement of 1~ns. The TARGET 5 ASICs combine sampling, digitisation and analogue triggering in the same package. Coupling between sampling and triggering operations limit the CHEC-M trigger sensitivity to $\sim$5~pe per pixel depending on the gain of the pixel in question. Simulations indicate that the desired trigger threshold per analogue of four pixels is $\sim$10~pe, corresponding to a minimum of $\sim$2.5~pe per pixel. 

\paragraph{On-Telescope Testing}

A prototype GCT telescope structure at the Observatoire de Paris in Meudon near Paris has been used for CHEC-M field tests over two intensive observing campaigns. The first Cherenkov light seen by a CTA prototype, and a dual-mirror telescope, was recorded during the first campaign in November 2015~\cite{jason-gamma16}.

\begin{figure}[t]
	\centering
	\resizebox{1\columnwidth}{!}{\includegraphics[trim=4.5cm 1.5cm 2cm 6cm, clip=true]{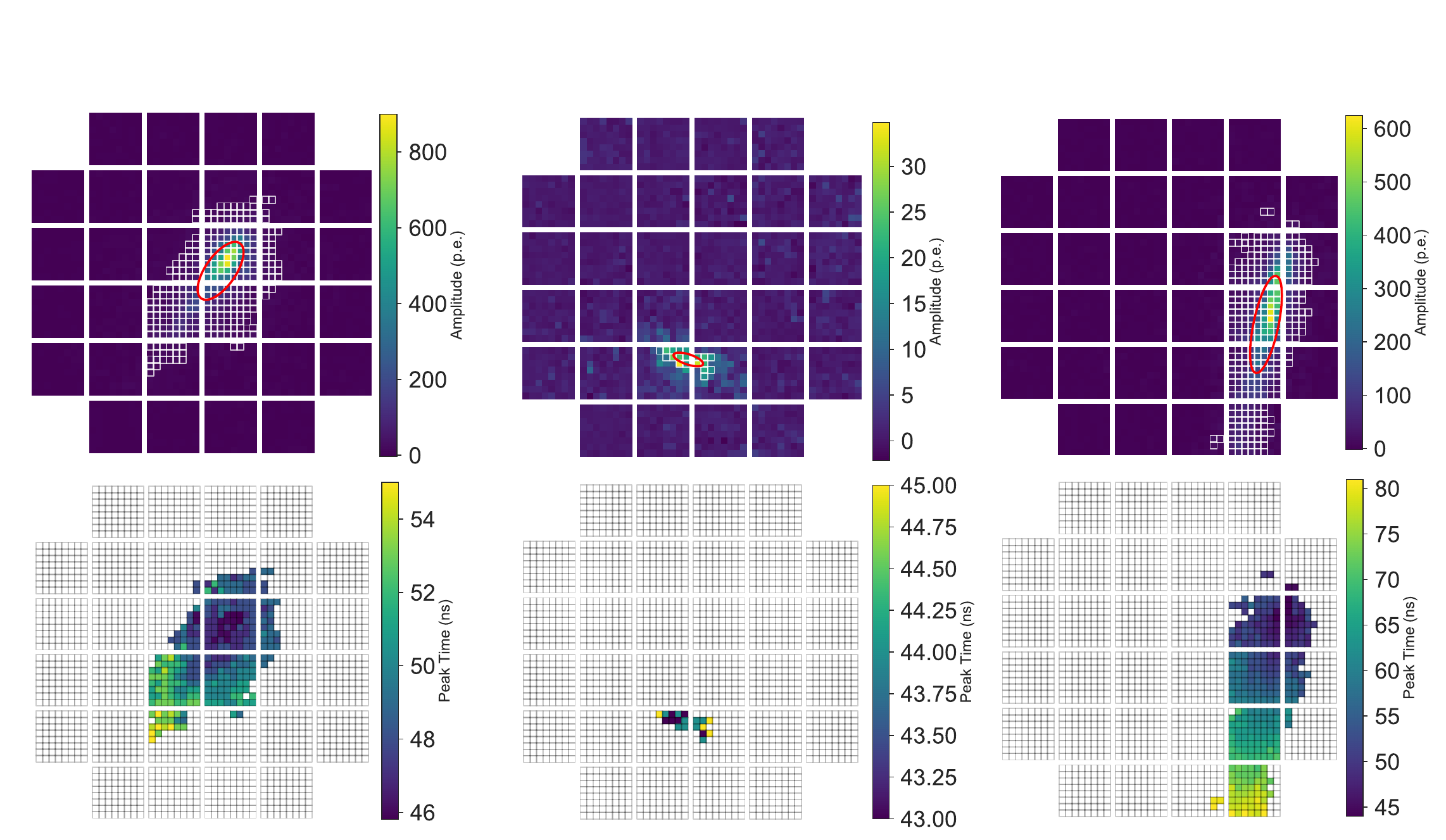}}
	\caption{Examples of three Cherenkov images recorded with CHEC-M on the GCT telescope prototype in Meudon, Paris. The intensity in each pixel is shown in the upper plots, whilst the peak arrival time per pixel is shown for the same images in the lower plots. For further explanation refer to the text.}
	\label{fig:res1}
\end{figure}

The Meudon observations were subject to NSB levels roughly 20 to 100 times brighter than that anticipated on the final CTA site. This required CHEC-M to be operated at a low gain, far from the 1100~V point at which the previously mentioned absolute gain calibration is possible, resulting in an RMS spread in the calibrated gain between all pixels of $\sim$15\%. Figure~\ref{fig:res1} shows three examples of on-sky Cherenkov images from the second observing campaign in Spring 2017. The upper most images indicate the calibrated image intensity in pe for each camera pixel. Following calibration the images are cleaned and the Hillas parameters~\cite{hillas} of the remaining image pixels are extracted. The white boxes outlining part of the images indicate pixels that survive image cleaning. In this step, all pixels containing a signal greater than 20~pe and at least one neighbouring pixel with a signal larger than 10~pe are retained. Ellipses resulting from the extracted Hillas parameters are then shown in red. A more detailed explanation of image analysis and a comparison to Monte Carlo simulations can be found in~\cite{gct}. The lower images in Figure~\ref{fig:res1} show the arrival time of the peak of the Cherenkov light flash in each pixel across the camera for the same events as the upper images. The images can be seen to propagate across the focal plane in time, as expected for a Cherenkov flash from a shower inclined with respect to the telescope focal plane. This additional time information is only possible due to the waveform sampling nature of the camera electronics, and will be useful for advanced image cleaning, background rejection and event reconstruction algorithms. Images at the highest energies can take many tens of nanoseconds to propagate across the camera - see for example the right-most Cherenkov image in Figure~\ref{fig:res1}. Without a $\sim$100~ns read out window such images would appear truncated, with the attendant negative impact on image analysis. All images recorded on-site by CHEC-M can be attributed to air-showers stemming from cosmic-rays rather than gamma rays. 

\begin{figure}[t]
	\centering
	\resizebox{1\columnwidth}{!}{\includegraphics[trim=2cm 2cm 2cm 7cm, clip=true]{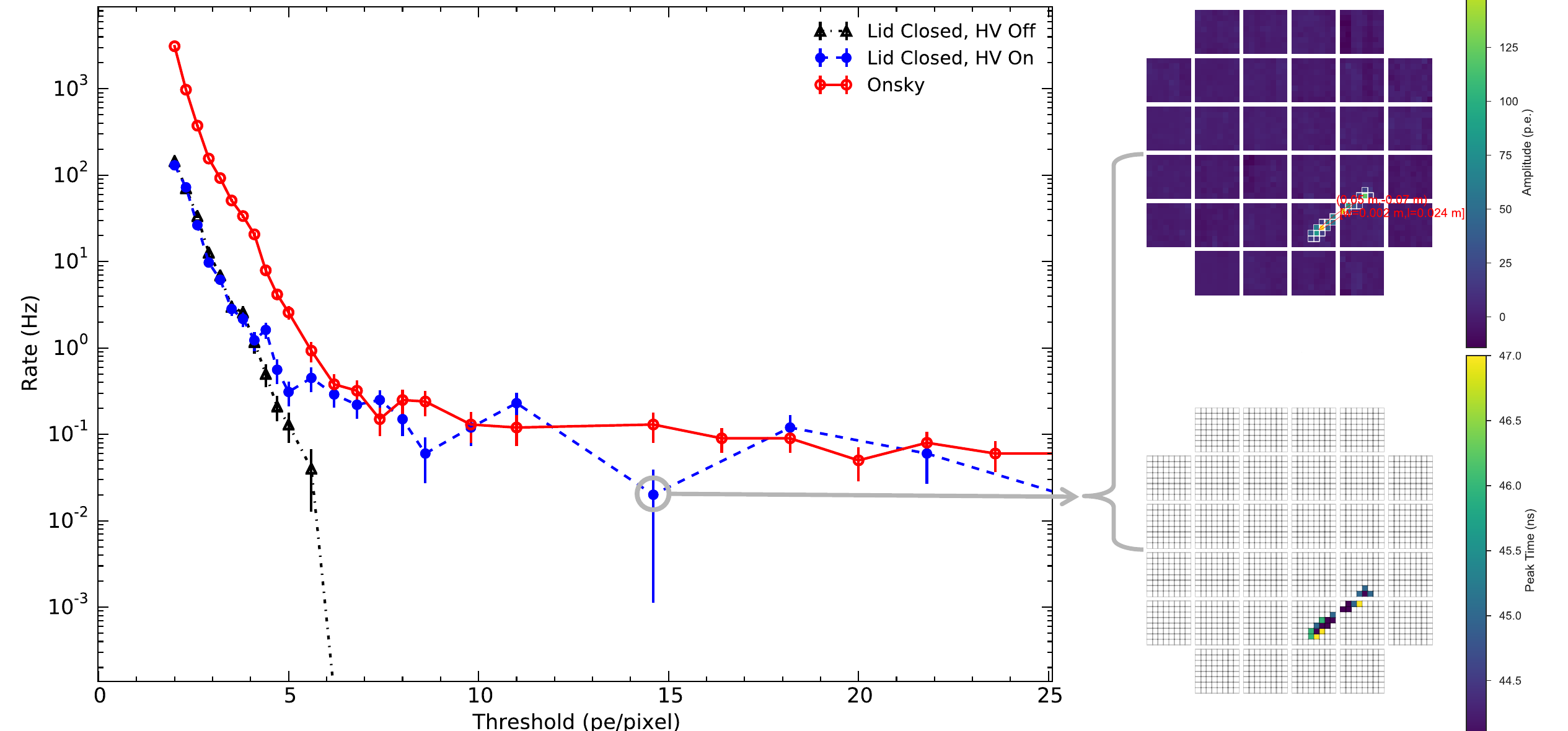}}
	\caption{Trigger rate as a function of camera-trigger threshold as recorded for CHEC-M on the GCT telescope prototype in Meudon, Paris. The rate recorded with the camera lid closed and the MAPM HV off is indicated by the black triangles (broken-dashed line). The rate with the MAPM HV on is shown by the blue filled circles (dashed line). The rate recorded on-sky at an elevation angle of 60$^\circ$ is then shown by the red open circles (solid line). Inset to the right an image recorded with the HV on and the lid closed is shown (intensity in each pixel to the top and peak arrival time per pixel to the bottom).}
	\label{fig:res2}
\end{figure}

Scans of camera trigger rate as a function of threshold were taken on-site to determine an appropriate operating threshold as shown in Figure~\ref{fig:res2}. The TARGET 5 sampling was disabled to avoid the previously mentioned problem of coupling between sampling and triggering operations during the particular scan shown in this figure. Data taken with the MAPM HV off show the level of electronic noise in the system (black points). Turning the HV on with the camera lid closed results in the blue points. Here, a steady rate of events at the 0.1~Hz level are observed in the telescope park position (0$^\circ$ elevation) above a threshold of roughly 5~pe. This rate is attributed to cosmic rays interacting directly with the camera. An example of such an event is shown inset to the right of Figure~\ref{fig:res2}. On-sky data with the telescope at 60$^\circ$ elevation is shown by the red data points. In this data set, accidental triggers due to fluctuations in the NSB dominate the trigger rate of the camera when operating at low thresholds. Above a threshold of approximately 5~pe the rate is dominated by triggers from Cherenkov light initiated by cosmic-rays, examples of which can be seen in the previously discussed Figure~\ref{fig:res1}. Roughly 10$\%$ of these triggers were characteristic of cosmic rays interacting directly with the camera. The reduction in rate of cosmic rays interacting directly with the camera on-sky (red points) from that seen with the lid closed (blue points) is due to the geometric effect of elevating the camera. The unique geometry and fast time profile of these events make them easy to isolate. As a result of these trigger investigations the camera trigger threshold was set to $\approx$11~pe pixel resulting in a steady Cherenkov event rate of $\approx$0.1~Hz during normal operations in Meudon.  

\paragraph{Lessons Learnt}

The goal of the CHEC project is to provide a reliable, high-performance, product for CTA. The prototyping of CHEC-M has proven an invaluable step in this development path. Deployment on a functioning telescope structure helped to verify interfaces and to improve the planned operations procedure. Regular operation has proven useful in understanding system stability and reliability. On-sky data is proving useful in the on-going development of the data-analysis chain and in understanding the levels of calibration that will be required for CTA. The CHEC-M BEE were a success. Triggering, clock-synchronisation and data-transmission from 2~k channels of electronics all worked well, and the BEE for the final production-phase system will be very similar to that used in the prototype. However, CHEC-M does not meet all of the CTA performance requirements. The trigger noise incurred with sampling enabled and the non-uniformity in gain being of greatest concern.

\subsection{CHEC-S}
\label{checs}

CHEC-S integration and testing will take place this year. Component commissioning and testing is currently underway. By design, CHEC-S will tackle the limiting factors in the CHEC-M performance. The main differences between CHEC-M and CHEC-S are the use of SiPMs and an upgrade of the TARGET ASIC. 

SiPMs will allow gain measurements more easily for a range of input illumination levels and, critically, bias voltages. The gain may even be determined in the absence of light from the dark counts intrinsic to the SiPM. Improved knowledge of the gain will in turn improve the off-line calibration and charge reconstruction. Simulations show that an RMS spread of better than 10\% between all pixels is needed to meet CTA requirements. The SiPM gain spread is also intrinsically much less than that seen in MAPMs, and the bias voltage is adjustable per four camera pixels, so gain matching to much higher precision than in CHEC-M will be possible. However, the use of SiPMs is not without consequence. SiPMs are also subject to optical cross talk and detection efficiency variation as a function of bias voltage. The gain of SiPMs is also temperature sensitive, and for CHEC-S will drop by approximately 10-20\% over a 10$^\circ$~C temperature increase. This must be corrected for, either by temperature stabilisation, online adjustment of the bias voltage or offline correction via constant gain monitoring or knowledge of the temperature dependence. To this end, the liquid cooled focal plane plate will stabilise the temperature to within $\pm$1$^\circ$~C over time scales for which the gain may easily be re-measured in-situ. Due to their physical dimensions, the SiPM tiles result in less dead space in the focal plane than MAPMs. They also exhibit a more uniform angular response and higher detection efficiency (approximately 20\% higher). The SiPMs have a significantly different wavelength response to the MAPMs. The resulting increase in the NSB rate can be mitigated using optical filters, and the SiPMs provide significantly improved efficiency for signal photons. Dark count rates from the SiPMs at the nominal operating gain and temperature have been measured to be less than $\sim20\%$ of the expected dark sky NSB rate, ensuring a negligible impact on performance. Due to the undesirable coupling between sampling and triggering in the TARGET 5 ASICs, functionalities were split into two separate ASICs. T5TEA provides triggering based on the same concept as TARGET 5, with a sensitivity reaching the single pe level and a trigger noise of 0.25~pe for the CHEC-S gain. TARGET C performs sampling and digitisation, with a $\sim$70\% larger dynamic range and with an improvement in charge resolution by a factor $>$2 with respect to TARGET 5~\cite{target-gamma16}. 

\section{Future Prospects}
\label{future}

Plans are underway following prototyping to build, test and deploy three "pre-production" CHEC cameras on the Southern-Hemisphere CTA site. Beyond this, we aim to provide cameras for a significant fraction of the 70 baseline SSTs during the production phase of CTA. With the notable exception of the SiPMs it is expected that the majority of components used in CHEC-S will also be used in the final production design of CHEC. The Hamamatsu S12642 SiPM tiles used in CHEC-S will meet most, if not all of the CTA requirements. However, SiPM technology is rapidly evolving and the latest devices offer significant performance improvements, including increased photo-detection efficiency, lower optical cross talk and a reduced dependency of gain on temperature~\cite{sipm, sipm-nepomuk}. The camera nominally uses 2048 6~mm pixels. The use of slightly larger pixels up to 7~mm may result in improved performance due to the enlarged field of view despite an increase in NSB, cost (which both scale as pixel area) and reduced angular resolution. Simulations to explore this trade off, and laboratory tests of the latest SiPMs and simulations with different pixel sizes are ongoing, with the aim of choosing a photosensor for the pre-production CHEC cameras in the coming months. 

\acknowledgments
\label{ack}
We gratefully acknowledge financial support from the agencies and organizations listed here: http://www.cta-observatory.org/consortium\_acknowledgments.

\end{document}